\documentclass[prl,twocolumn,superscriptaddress]{revtex4}
\usepackage{amsfonts}
\usepackage{amsmath}    
\usepackage[normalem]{ulem}
\usepackage{bm}
\usepackage{graphicx}
\usepackage{natbib}

\renewcommand{\eqref}[1]{Eq.~(\ref{#1})}

\DeclareMathOperator{\sech}{sech}

\begin{document}
\title{Efficient and long-lived Zeeman-sublevel atomic population storage in an erbium-doped glass fiber}
\author{Erhan Saglamyurek }
\thanks{E.S., T.L., and L.V. contributed equally to this work.}
\affiliation{Institute for Quantum Science and Technology, and Department of Physics \& Astronomy, University of Calgary, 2500 University Drive NW, Calgary, Alberta T2N 1N4, Canada}

\author{Thomas Lutz }
\thanks{E.S., T.L., and L.V. contributed equally to this work.}
\affiliation{Institute for Quantum Science and Technology, and Department of Physics \& Astronomy, University of Calgary, 2500 University Drive NW, Calgary, Alberta T2N 1N4, Canada}

\author{Lucile Veissier}
\thanks{E.S., T.L., and L.V. contributed equally to this work.}
\affiliation{Institute for Quantum Science and Technology, and Department of Physics \& Astronomy, University of Calgary, 2500 University Drive NW, Calgary, Alberta T2N 1N4, Canada}
\author{Morgan P. Hedges}
\altaffiliation{Present address: Physics Department, Princeton University, Princeton, New Jersey, 08544, USA}
\affiliation{Institute for Quantum Science and Technology, and Department of Physics \& Astronomy, University of Calgary, 2500 University Drive NW, Calgary, Alberta T2N 1N4, Canada}
\author{Charles W. Thiel}
\affiliation{Department of Physics, Montana State University, Bozeman, Montana 59717, USA}
\author{Rufus L. Cone}
\affiliation{Department of Physics, Montana State University, Bozeman, Montana 59717, USA}
\author{Wolfgang~Tittel}
\email{wtittel@ucalgary.ca}
\affiliation{Institute for Quantum Science and Technology, and Department of Physics \& Astronomy, University of Calgary, 2500 University Drive NW, Calgary, Alberta T2N 1N4, Canada}


\begin{abstract}
Long-lived population storage in optically pumped levels of rare-earth ions doped into solids, referred to as persistent spectral hole burning, is of significant fundamental and technological interest. However, the demonstration of deep and persistent holes in rare-earth ion doped amorphous hosts, e.g. glasses, has remained an open challenge since many decades -- a fact that motivates our work towards a better understanding of the interaction between impurities and vibrational modes in glasses. Here we report the first observation and detailed characterization of such holes in an erbium-doped silica glass fiber cooled to below 1 K. We demonstrate population storage in electronic Zeeman-sublevels of the erbium ground state with lifetimes up to 30 seconds and 80\% spin polarization. In addition to its fundamental aspect, our investigation reveals a potential technological application of rare-earth ion doped amorphous materials, including at telecommunication wavelength. 
\end{abstract}

\maketitle
\newpage

Persistent spectral hole burning in cryogenically cooled rare-earth ion (REI) doped solids, the process of transferring atomic population via optical pumping from the ground level of the REI into a long-lived (metastable) level, has been extensively studied for five decades \cite{macfarlane_coherent_1987,sun_rare_2005}. The observation of persistent holes is of fundamental interest as they allow characterizing the interaction between REIs and their environment, and they find applications in both classical and quantum information processing \cite{nilsson_initial_2002}. Deep, narrow, and long-lived holes have been observed for various REIs doped into crystalline hosts \cite{macfarlane_spectral_1987,hastings-simon_spectral_2008,konz_temperature_2003}. These holes generally result from population storage in nuclear spin hyperfine levels, or, for REIs having an unpaired electron spin (so-called Kramers ions), in electronic Zeeman sublevels. Compared to REI-doped crystals, REIs in glasses experience much larger inhomogeneity of optical and spin transitions, as well as coupling with inelastic tunneling modes (also referred-to as two-level systems -- TLS) \cite{selzer_anomalous_1976, broer_low-temperature_1986, littau_dynamics_1992} that results in significantly different decoherence and population relaxation properties. Persistent population transfer due to coupling with TLS, photochemical hole burning, and using spin states have been reported more than two decades ago \cite{hayes_mechanisms_1978,jankowiak_spectral_1993,Schmidt1994}, but the interaction between REIs and vibrational modes in amorphous media -- the generalization of the well-known phonons that are present in crystals -- is still not well understood. Furthermore, the observed holes have been shallow or hundreds of MHz wide, making them unlikely to be suitable for information processing applications.

An exception is the erbium-doped fiber that we have used recently for quantum state storage \cite{Saglamyurek_2015, Jin_2015}. The goal of the present paper is to detail its hole-burning mechanism. We observe deep, persistent holes with 15 MHz width (limited by power-broadening) and lifetimes approaching a minute that arise from population redistribution among the Zeeman-split electronic sublevels of the Er$^{3+}$ ground state. This improves upon the properties observed before for REI-doped glasses in terms of spin polarization (i.e. hole depth) and spectral width \cite{MacFarlane1987, Schmidt1994}. We furthermore characterize the dependence of the spin relaxation rate on magnetic field, temperature, wavelength, and erbium ion concentration. This allows us to determine the processes responsible for electron spin state relaxation in this amorphous medium and provides the necessary information to optimize operation parameters for practical applications.
 
Except where mentioned otherwise, our experiments employ a commercially available, 20 m-long silica fiber co-doped with 190 ppm erbium (INO S/N 404-28252). The fiber is cooled to temperatures $T$ as low as 0.65 K using an adiabatic demagnetization refrigerator, and a magnetic field $B$ of up to 0.25 T that lifts the Kramers degeneracy of the ground and excited electronic levels, splitting each into two Zeeman sublevels.  We burn spectral holes into the erbium ions' inhomogeneously broadened $^4I_{15/2}\leftrightarrow ^4I_{13/2}$  zero phonon line using 500 ms long laser pulses with a peak power of 1-5~$\mu$W derived from a narrowband, intensity-modulated, continuous wave laser operating, except where otherwise specified, at 1532 nm wavelength. After a variable waiting time that exceeds the 11 ms excited state lifetime, we probe the absorption profile by linearly chirped the frequency of the laser light over 500 MHz across the generated hole. The 1 ms-long scans are implemented using a phase modulator driven by a serrodyne signal \cite{Johnson1988}. Depending on the experimental conditions, we observe persistent spectral holes with line widths as narrow as 15 MHz, including the effect of power broadening. However, from coherence lifetime measurements, we know that homogeneous linewidths as narrow as a few MHz can be achieved in our fiber \cite{in-prep}. By measuring the area of the spectral hole with respect to the waiting time, we characterize the decay dynamics of the spectral hole and extract the underlying population relaxation processes.  

\begin{figure}[t!]
\centering 
\includegraphics[width=0.9\columnwidth]{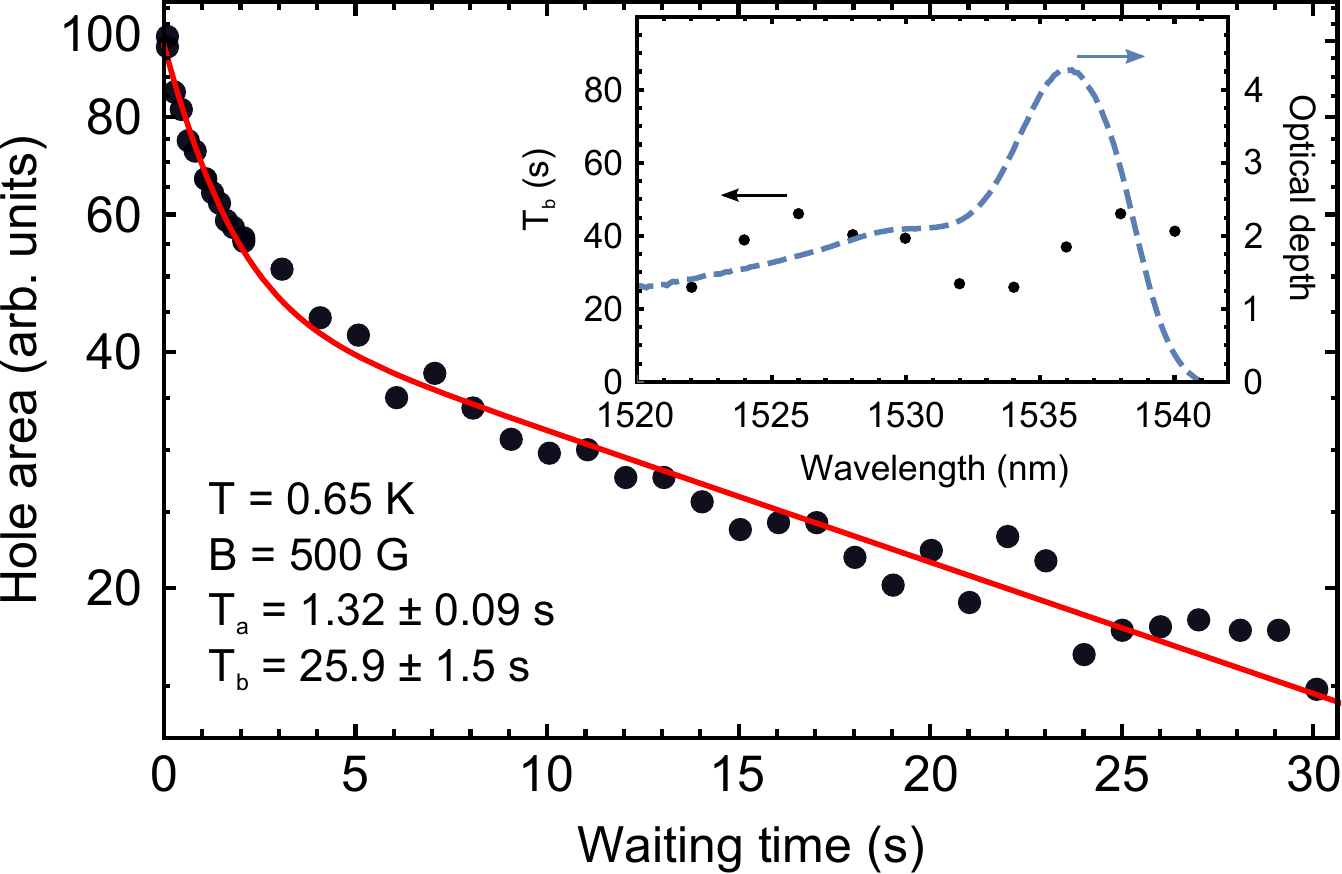}
\caption{Spectral hole area as a function of the waiting time for $B=500$~G and $T= 0.65$~K. The experimental data is fitted by the sum of two exponential decays with characteristic times $T_a = 1.32 \pm 0.09$~s and $T_b = 25.9 \pm 1.5$~s. The inset shows the fiber absorption profile at $T=0.8$~K (dashed blue line) and the hole lifetime $T_b$ for various wavelengths (black dots).}
\label{fig:decay_wdep}
\end{figure}

First, we determine the decay dynamics for persistent holes, with a typical example 
shown in Fig.~\ref{fig:decay_wdep}. We find that all decays, for temperatures ranging from 0.65 to 3.5 K and magnetic fields from 0 to 0.25 T, can be described by two exponential functions with similar weights and decay times ($T_a$) on the order of a second and ($T_b$) on the order of a few tens of seconds. This observation could indicate the presence of two different classes of erbium ions in the fiber, or could result from the average over a non-uniform, continuous distribution of decays corresponding to ions with different spin-state lifetimes. In addition, we characterize the wavelength dependence of the persistent hole lifetimes $T_b$ at a temperature of 0.8 K; see inset of Fig.~\ref{fig:decay_wdep}. We observe long-lived spectral holes with comparable lifetimes across the entire THz-wide inhomogenous absorption line.
 
Second, we assess the efficiency of the hole burning mechanism, which we define as the maximum degree of spin polarization. It characterizes how deep a persistent spectral hole can be burnt. Spectral holes with absorption reduced by as much as 80\% compared to the optical depth of the erbium-doped fiber before spectral hole burning were observed at $T= 0.7$~K, and with a probing delay of 50 ms. Furthermore, as the temperature rises from 0.7 to 3.5~K, the hole depth decreases linearly. In consequence, we extrapolate that it is, at least in principle, possible to burn a 100\% deep hole at an optimal temperature of 0 K. This fact alone excludes population redistribution via the TLS mechanism, as the latter generally only allows for persistent holes with a maximum depth of 50\%, and typically much less --  even at $T=0$~K \cite{hayes_non-photochemical_1978}. Furthermore, given the width of the observed holes and the small intensity used to create them, we can also exclude photochemical hole burning.  

\begin{figure}[t]
\includegraphics[width=0.9\columnwidth]{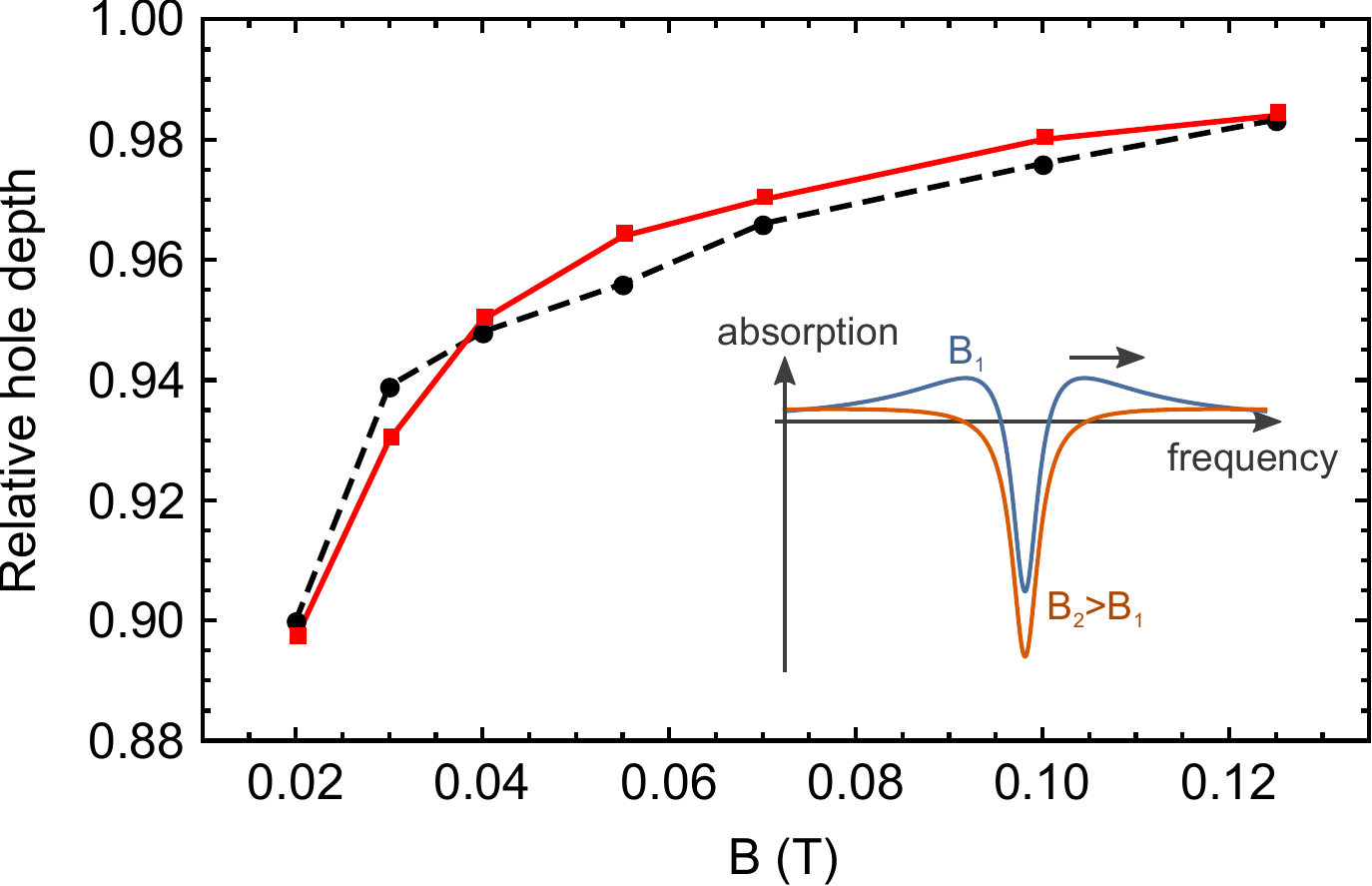}
\caption{Magnetic field dependent depth of a 200~MHz-wide hole normalized to the depths of a 100 MHz-wide hole at each field, which is not significantly affected by population transfer from an anti-hole. Please note that the absolute hole depth of a narrow hole decreases at $B > 0.06$~T due to decreasing lifetimes, as shown in Fig.~\ref{fig:conc-dep}~(c). Experimental data is shown by black dots; simulated results are in red.}
\label{fig:zeeman2}
\end{figure}

An unambiguous way to confirm our hypothesis of Zeeman level storage as the hole burning mechanism would be to burn one narrow spectral hole, observe the regions of increased absorption due to population redistribution, i.e. anti-holes, and extract their magnetic field-dependent shift. However, this was not possible in our case due to the disorder in the amorphous material, which leads to very broad anti-holes that cannot be resolved. To indirectly verify the presence and positions of these anti-holes, we employ a novel method that is based on burning a wide hole and measuring its depth as a function of magnetic field. The results for a 200 MHz-wide hole are shown in Fig.~\ref{fig:zeeman2}. As the field decreases, the anti-holes begin to overlap with the central hole, causing it to become more shallow (due to the first-order electronic Zeeman splitting, the frequency shift between the hole and anti-hole depends linearly on the magnetic field strength). Modeling the hole and the anti-holes with Lorentzians of magnetic field-dependent width and position, we can predict the change in hole-depth as a function of the magnetic field, see Fig.~\ref{fig:zeeman2}. This allows estimating a mean hole/anti-hole splitting of 25 GHz/T, which is comparable to the values typically measured in erbium doped crystals \cite{bottger_spectroscopy_2006,Hastings2008}. This further supports our conclusion of population storage in Zeeman levels. We also find that the anti-hole width is about six times larger than the splitting between hole and anti-hole, which reflects the large inhomogenous broadening of the Zeeman transition in the fiber, as discussed earlier. 

\begin{figure}[t]
\centering
\includegraphics[width=0.9\columnwidth]{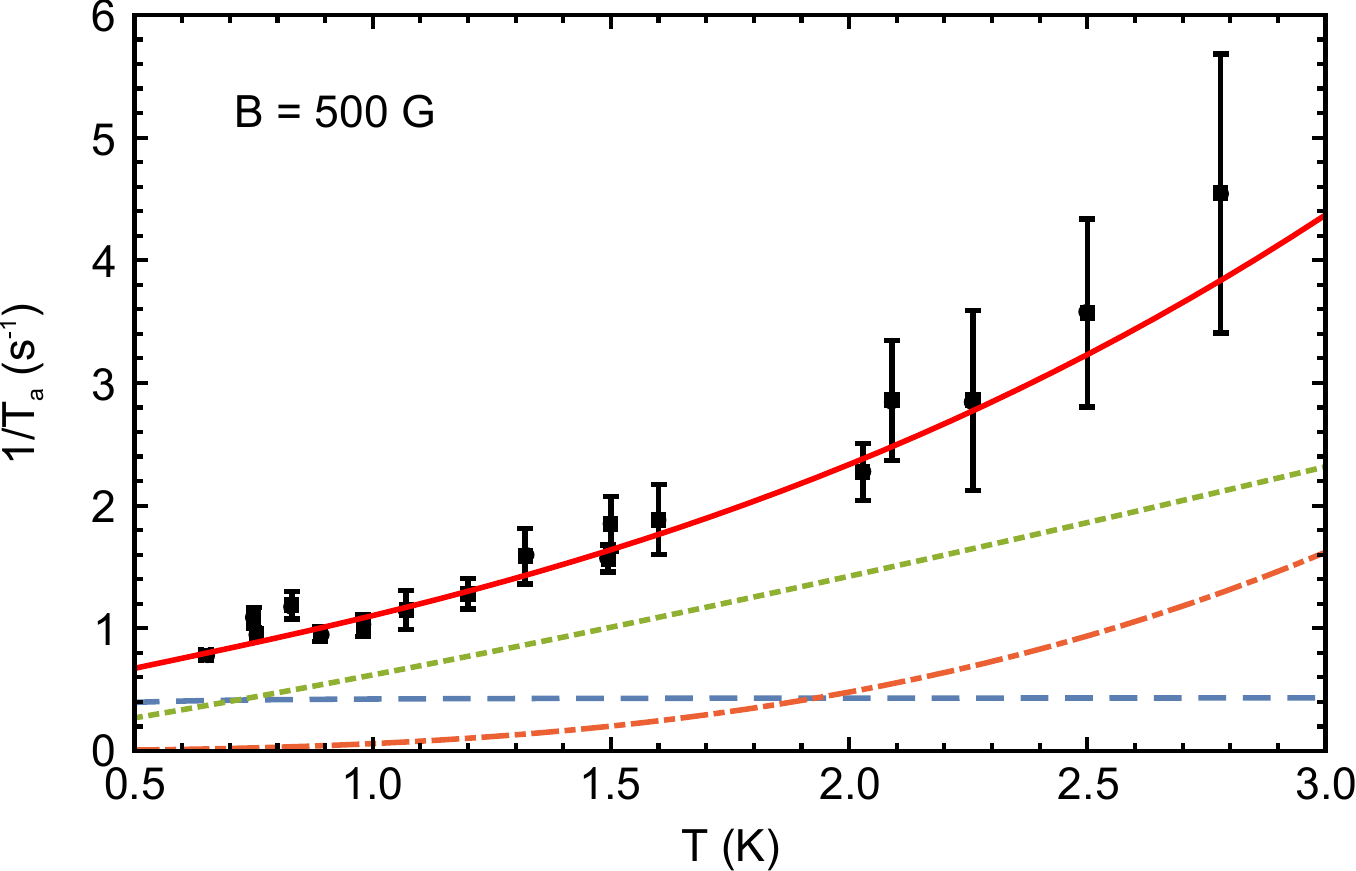}
\caption{Spin relaxation rate $1/T_a$ as a function of temperature at $B=500$~G. The dashed lines correspond to the different terms of Eq.~\ref{eq:model}: the Er-Er coupling (blue dashed), the direct process (green dotted), and the Raman process (orange dashed dotted).}
\label{fig:t-dep}
\end{figure}

\begin{figure}[t]
\includegraphics[width=0.9\columnwidth]{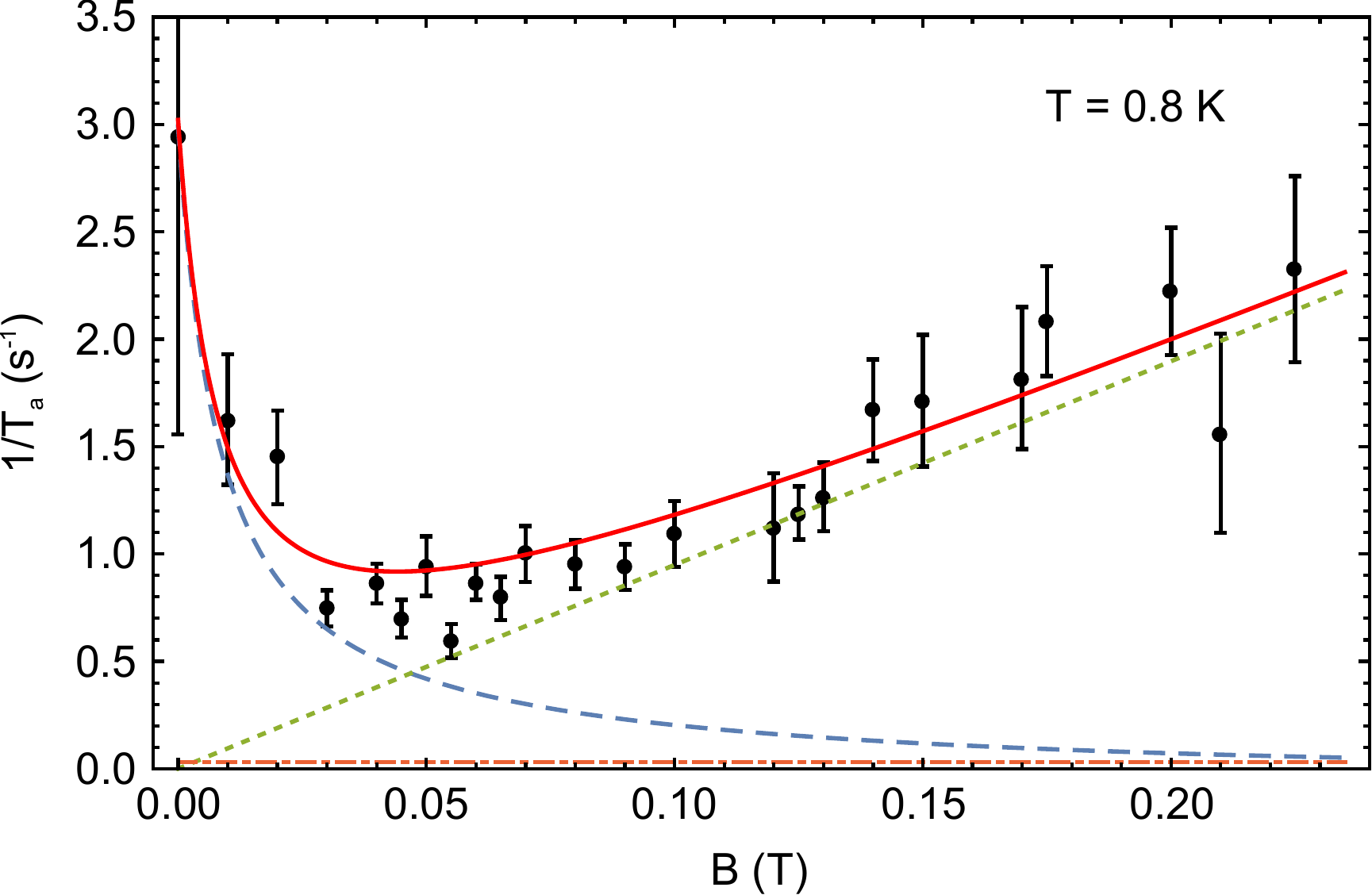}
\caption{Dependence of the rate $1/T_a$ on the magnetic field at $T=0.8$~K. The solid line shows the theoretical prediction of Eq.~\ref{eq:model} with the set of parameters in Table~\ref{table:parameters}, and the dashed and dotted lines correspond to the individual terms  (first term: blue dashed, second term: green dotted, third term: orange dashed dotted). }
\label{fig:b-dep}
\end{figure}

To gain insight into the relaxation mechanisms that determine the observed spin lifetimes as well as to identify optimal conditions for potential applications based on persistent spectral hole burning, we experimentally and theoretically study magnetic field and temperature dependence of hole decay rates (the inverse of the hole lifetime); examples of the experimental data are shown in Figs.  \ref{fig:t-dep} and \ref{fig:b-dep}. We note that the two decay components $T_{a}$ and $T_{b}$ exhibit very similar behavior as a function of magnetic field and temperature for the assessed range of parameters. Given the difficulty of measuring the longer Zeeman lifetime $T_b$ with high precision at high temperature, we consider only the fast decay component $T_a$ in the following analysis.
 
To describe our experimental data, we develop a model based on the framework proposed in \cite{Bottger2006} for an Er:Y$_2$SiO$_5$ crystal, but adapt it to an amorphous host, in which the Debye model of phonons is not applicable and has to be replaced by TLS and local vibrational modes. We assume the spin relaxation rate to be described by
\begin{equation}
\frac{1}{T_{a/b}} = \frac{\alpha_1}{\Gamma_S^0 + \gamma B} \sech ^2\left( \frac{g \mu_B B}{2 k T} \right) + \alpha_2 B^l T^m + \alpha_3 T^n \, .
\label{eq:model}
\end{equation}
\noindent
The first term corresponds to the average mutual spin flip-flop rate due to magnetic dipole-dipole interactions between erbium ions, with $g$ the $g$-factor of the Er ions, $\mu_B$ the Bohr magneton and $k$ the Boltzmann constant. We assume $g=9$, which is consistent with averaging over all possible direction-dependent values, ranging from 0 to 18 \cite{Kurkin1980,bottger_laser_2002,milori_optical_1995,ball_low-temperature_1961,Macfarlane2006,Staudt2006720}. 
Provided the erbium ions are uniformly distributed, the coefficient $\alpha_{\rm 1}$ scales quadratically with Er concentration (we will examine this assumption below). In addition, the spin flip-flop term includes broadening of the inhomogeneous linewidth $\Gamma_S$ of the spin transition with magnetic field, i.e. $\Gamma_S=\Gamma_S^0+\gamma B$. This also leads to the anti-hole broadening discussed in the previous analysis, from which we extracted $\gamma=6\times 25$~GHz/T.

The second term describes the direct coupling between Er ions and resonant, thermally-driven vibrational or TLS modes of the glass matrix. While this effect is in some ways analogous to the coupling with phonons in crystals, one cannot expect it to have the same dependence on the magnetic field and temperature due to a different density of states of vibrational modes in the amorphous medium \cite{buchenau_low-frequency_1986}. We therefore include free parameters $l$ and $m$ in Eq.~\ref{eq:model}. 

The third term describes the process where the Er ion relaxes through higher-energy vibrational or TLS modes of the glass via a second-order Raman-type interaction. It is equivalent to the phonon mediated inelastic Raman relaxation in crystals, which involves two off-resonant vibrational modes. In the case of a crystal doped with a Kramers ion such as erbium, the Raman process scales as $T^9$ \cite{Orbach1961}; however, since the density of states of vibrational modes might be very different in the fiber, the temperature dependence of the Raman process is described by the free parameter $n$. 

\begin{table}[t]
\begin{center}
\begin{tabular}{|c|c|c|c|}
\hline
$g$ & 9 & \multicolumn{1}{||c|}{$l$} & $1 \pm 0.1$ \\
$\gamma$ (GHz/T) & 150  & \multicolumn{1}{||c|}{ $m$} & $1.2 \pm 0.2$\\
$\Gamma_S^0$ (GHz) & $1.3\pm 0.2$  & \multicolumn{1}{||c|}{$n$} & $3 \pm 0.5$\\
\hline  \hline
1/$T_a$ & \multicolumn{1}{c}{$\alpha_1$} & \multicolumn{1}{c}{$\alpha_2$}  & $\alpha_3$  \\ \cline{1-4}
fiber 1  &  \multicolumn{1}{c}{$3.80 \pm 0.2$} &  \multicolumn{1}{c}{$12.4 \pm 0.5$}  &  \multicolumn{1}{c|}{$0.06 \pm 0.01$} \\
\hline \hline
$1/T_b$ & \multicolumn{1}{c}{fiber 1} & \multicolumn{1}{c}{fiber 2} & \multicolumn{1}{c|}{fiber 3} \\
\hline
$\alpha_1$ & \multicolumn{1}{c}{$0.160 \pm 0.2$} & \multicolumn{1}{c}{$0.234 \pm 0.2$} & \multicolumn{1}{c|}{$1.67  \pm 0.2$} \\
$\alpha_2$ & \multicolumn{1}{c}{$0.72 \pm 0.5$} & \multicolumn{1}{c}{$0.82 \pm 0.5$} &  \multicolumn{1}{c|}{$4.7 \pm 0.5$} \\
\hline
\end{tabular}
\end{center}
\caption{Parameters for Eq.~\ref{eq:model} to fit the experimental data, with $\alpha_1$ in $10^{9}$ s$^{-2}$, $\alpha_2$ in Hz T$^{-1}$ K$^{-1}$ and $\alpha_3$ in Hz T$^{-4}$.}
\label{table:parameters}
\end{table}

By iteratively fitting Eq.~\ref{eq:model} to several subsets of experimental data, we obtain the single set of parameters given in Table~\ref{table:parameters} that allows describing all experimental data. The good agreement is exemplified with the two datasets shown in Figs.~\ref{fig:b-dep} and \ref{fig:t-dep}. We note that the value for the inhomogeneous line width at zero field, $\Gamma_S^0$, exceeds the value for REI-doped crystals by several orders of magnitude \cite{fraval_dynamic_2005,probst_anisotropic_2013,jobez_coherent_2015}. While we have no explanation yet for why $l$=1, we note that it is not possible to attribute this dependence to a direct-phonon-type process as observed in crystals because such a process must scale at least quadratically \cite{Orbach1961} with the magnetic field. This argument is based on the assumption that the phonon density of states is not decreasing with frequency, which is correct for silica glass at our temperatures \cite{Haworth2010}. Furthermore, we find that the fitted value for $m$, describing the temperature-dependent part of the coupling to TLS, is consistent with values that allow modeling coherence properties in REI-doped glasses \cite{Schmidt1994,Macfarlane2006,sun_exceptionally_2006}, and that a difference of $n$ from the value in crystals (i.e. $n_{crystal}$=9) has also been observed in nanometer size particles \cite{meltzer_electron-phonon_2000}.  

Let us now have a look how the interplay of the three different processes in Eq.~\ref{eq:model} -- represented as individual lines in Figs.~\ref{fig:t-dep} and \ref{fig:b-dep} -- determines the lifetime of the spectral hole. Fig.~\ref{fig:t-dep} shows that the decay rate increases monotonically with temperature, and that, for higher temperature, it is dominated by the Raman-type interaction. Furthermore, Fig.~\ref{fig:b-dep} shows that spin flip-flops, i.e. the exchange of spin states between neighboring, resonant erbium ions through magnetic dipole-dipole interactions, dominate the decay rate at small magnetic fields and 0.8~K. However, the contribution of this process decreases rapidly as the field is increased due to the additional inhomogeneous broadening described by $\gamma B$, i.e. a decreasing probability for neighboring ions to be resonant. For higher fields, the decay rate is dominated by coupling to thermally-driven TLS and vibrational modes. Hence, for long-lived holes, it is always beneficial to work at the lowest possible temperature, but there is an optimum magnetic field. For our fiber, the best value for $T_a$ is obtained at 0.65~K and around 500~G.

\begin{figure}[t!]
\centering
\includegraphics[width=0.95\columnwidth]{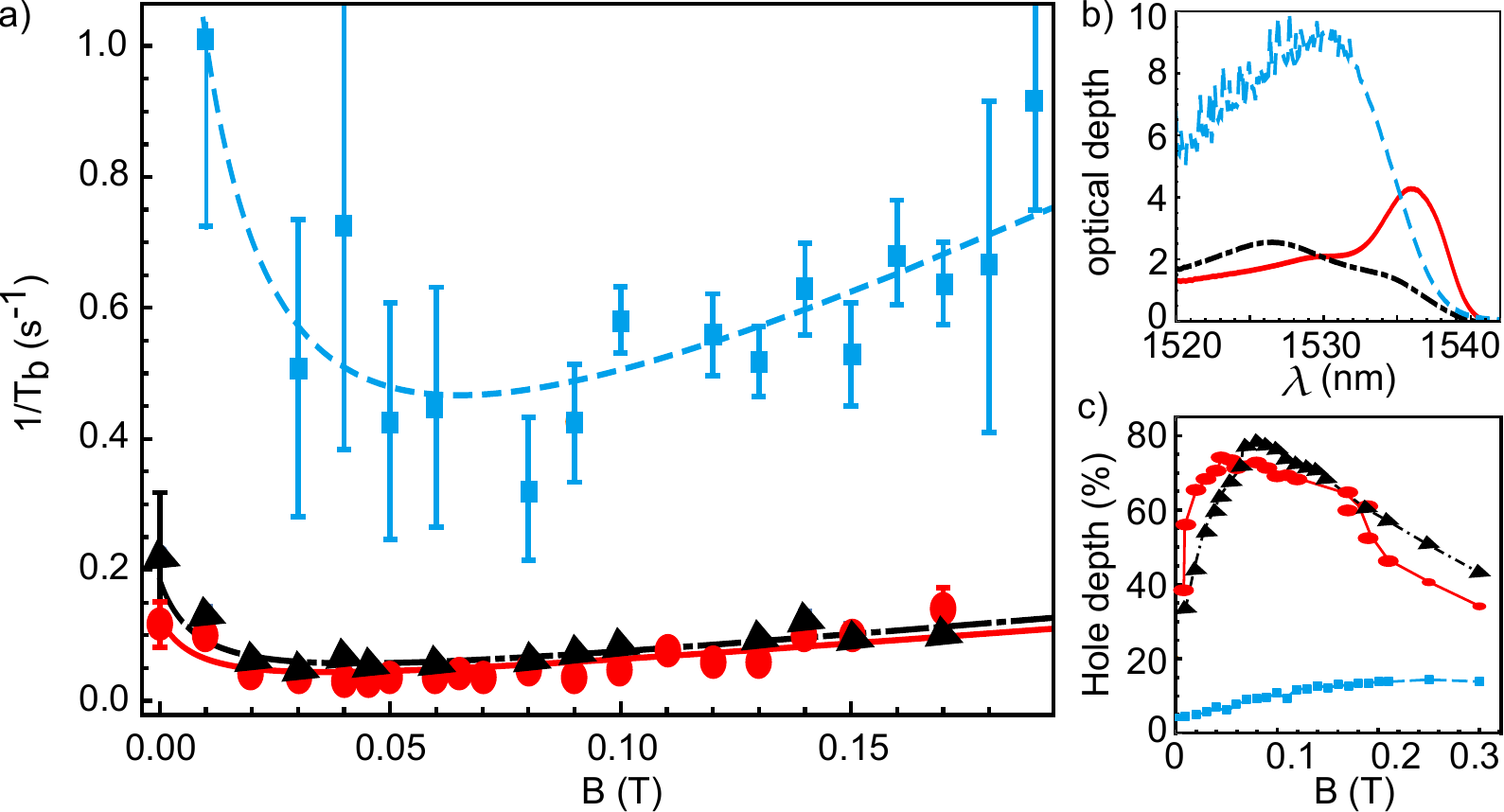}
\caption{(a) Dependence of the rate $1/T_b$ on the magnetic field for fibers with different erbium doping concentrations and lengths. (b) Absorption profiles of the three fibers. (c) Hole depth as a function of magnetic field. All measurements are taken at 0.8 K, at 1532~nm for fibers 1 and 2, and at 1536~nm for fiber 3. Red: fiber 1 (190 ppm, 20 m), black: fiber 2 (200 ppm, 10 m), blue: fiber 3 (1200 ppm, 3 m).}
\label{fig:conc-dep}
\end{figure}

Next, we characterize and compare the magnetic field dependence of the long decay rates $T_b$ at $T=0.8$~K for three fibers of different Er-doping concentration: fiber 1 (the same as in the investigations described above) -- 190 ppm (INO S/N 404-28252), fiber 2 -- 200 ppm (INO S/N 402-28254) and fiber 3 -- 1200 ppm (INO S/N 502-28255). Fig.~\ref{fig:conc-dep}~(b) shows the absorption profiles for the three fibers. The profiles for the two low-concentration fibers (fiber 1 and 2) differ greatly, which we attribute to a difference in the co-dopants, and those of fiber 2 and 3, which features a very different erbium concentration, are comparable, suggesting similar co-dopants. The results of the rate measurements are shown in Fig.~\ref{fig:conc-dep}~(a). We emphasize that the rates for fibers 1 and 2, which feature almost identical Er ion concentrations, are very similar over the entire range of magnetic field strength. Contrary to what one might expect, this seems to suggests that the spectral hole burning properties are not much affected by the co-dopants. However, comparing with the data obtained for fiber 3, we find that the decay rates strongly depend on the erbium doping concentration. To fit the data, we therefore include the concentration dependence only into $\alpha_1, \alpha_2$ and $\alpha_3$, and we assume that the spin inhomogeneous broadening, characterized by $\Gamma_S^0$ and $ \gamma$, is the same for the three fibers. 

As shown in Table~\ref{table:parameters}, we find that the value for $\alpha_1$ increases with Er doping concentration, which is consistent with the assumption that the first term in Eq.~\ref{eq:model}  describes spin flip flops. The concentration dependence of $\alpha_2$ suggests that the dominant relaxation mechanism at high magnetic fields is through spin-elastic TLS involving motion of Er ions rather than simple vibrational modes of glass -- or TLS modes created by Er ions \cite{Macfarlane2006}. For conclusive information about the scaling of these parameters, more investigations using fibers with other doping concentrations are needed. Furthermore, we find that the third term does not contribute to the decay rates measured at 0.8~K. 


Finally, we assess the quality of the spectral hole burning in terms of the hole depth for these three fibers after 50~ms waiting time and at $T=0.8$~K -- the results are shown in Fig.~\ref{fig:conc-dep}~(c). We find that the erbium ion concentration has a direct impact on the hole depth -- the smaller the concentration, the deeper the hole, which is due to reduced spin flip flops. Furthermore, we see that the optimum magnetic field also depends on the doping concentration, which is a consequence of the change in relative importance of the spin flip-flop process versus coupling with TLS. 

In conclusion, we established the existence of long-lived, narrow and deep persistent spectral holes generated via spin-level storage in a rare-earth-ion doped amorphous host material. More precisely, using a weakly doped silica fiber under an optimized magnetic field, we observed population storage in electronic Zeeman levels of erbium ions with lifetimes approaching a minute. We furthermore developed a model that identifies lifetime-limiting mechanisms, in particular spin flip-flops at low magnetic fields and coupling to TLS and vibrational modes at high fields. The model shows that the use of an amorphous host, which results in large spin inhomogeneous broadening, and low doping concentration leads to small spin flip-flop rates even at low magnetic fields, and hence to the possibility of observing deep and long-lived spectral holes. Our study shines light on the fundamental interaction between impurities and vibrational modes in glasses, which, as opposed to crystals, is still not well understood. Furthermore, our findings allow parameter optimization in future applications of persistent spectral hole burning, including optical quantum memories, photonic processors, configurable filters, and long-lived optical storage elements for fiber-optic communication. We emphasize that our investigations have been carried out on the 1532~nm transition in erbium, which opens these applications to the convenient telecommunication c-band.
\\

The authors thank Daniel Oblak, Neil Sinclair, and Roger Macfarlane for discussions, and acknowledge support from Alberta Innovates Technology Futures (ATIF), the National Engineering and Research Council of Canada (NSERC), and the National Science Foundation of the USA (NSF) under award nos. CHE-1416454 and PHY-1415628. W.T. is a senior fellow of the Canadian Institute for Advanced Research (CIFAR).


\begin{thebibliography}{10}
\newcommand{\enquote}[1]{``#1''}

\bibitem{macfarlane_coherent_1987}
R.~M. Macfarlane and R.~M. Shelby, in \emph{Modern {Problems} in {Condensed} {Matter} {Sciences}}, vol.~21 of \emph{Spectroscopy of {Solids} {Containing} {Rare} {Earth} {Ions}}, A.~A. Kaplyanskii and R.~M. Macfarlane, eds.
  (Elsevier, 1987), pp. 51--184.

\bibitem{sun_rare_2005}
Y.~C. Sun, in \emph{Spectroscopic {Properties} of {Rare} {Earths} in {Optical} {Materials}}, P.~R. Hull, P.~J. Parisi, P.~R. M.~O. Jr, P.~H. Warlimont, D.~G. Liu, and P.~B. Jacquier, eds. (Springer Berlin Heidelberg, 2005), no.~83 in Springer {Series} in {Materials} {Science}, pp. 379--429.

\bibitem{nilsson_initial_2002}
M.~Nilsson, L.~Rippe, N.~Ohlsson, T.~Christiansson, and S.~Kr\"{o}ll, Physica Scripta \textbf{T102}, 178 (2002).

\bibitem{macfarlane_spectral_1987}
R.~M. Macfarlane and J.~C. Vial, Physical Review B \textbf{36}, 3511 (1987).

\bibitem{hastings-simon_spectral_2008}
S.~R. Hastings-Simon, M.~Afzelius, J.~Minar, M.~U. Staudt, B.~Lauritzen, H.~de~Riedmatten, N.~Gisin, A.~Amari, A.~Walther, S.~Kr\"{o}ll, E.~Cavalli, and M.~Bettinelli, Physical Review B \textbf{77}, 125111 (2008).

\bibitem{konz_temperature_2003}
F.~K\"{o}nz, Y.~Sun, C.~W. Thiel, R.~L. Cone, R.~W. Equall, R.~L. Hutcheson, and R.~M. Macfarlane, Physical Review B \textbf{68}, 085109 (2003).

\bibitem{selzer_anomalous_1976}
P.~M. Selzer, D.~L. Huber, D.~S. Hamilton, W.~M. Yen, and M.~J. Weber, Physical Review Letters \textbf{36}, 813 (1976).

\bibitem{broer_low-temperature_1986}
M.~M. Broer, B.~Golding, W.~H. Haemmerle, J.~R. Simpson, and D.~L. Huber, Physical Review B \textbf{33}, 4160 (1986).

\bibitem{littau_dynamics_1992}
K.~A. Littau, M.~A. Dugan, S.~Chen, and M.~D. Fayer, The Journal of Chemical Physics \textbf{96}, 3484 (1992).
  
\bibitem{hayes_mechanisms_1978}
J.~M.~Hayes, and G.~~J.~Small, Chemical Physics Letters \textbf{54}, 435-438 (1978).

\bibitem{jankowiak_spectral_1993}
R.~Jankowiak, J.~M. Hayes, and G.~J. Small, Chemical Reviews \textbf{93}, 1471 (1993).

\bibitem{Schmidt1994}
T.~Schmidt, R.~M. Macfarlane, and S.~V\"{o}lker, Physical Review B \textbf{50}, 15707 (1994).

\bibitem{Saglamyurek_2015}
E.~Saglamyurek, J.~Jin, V.~B.~Verma, M.~D.~Shaw, F.~Marsili, S.~W.~Nam, , D.~Oblak, W.~Tittel. Nature Photonics \textbf{9}, 83-87 (2015).

\bibitem{Jin_2015} 
 J.~Jin, E.~Saglamyurek, M.~Grimau Puigibert, V.~B.~Verma, F.~Marsili, S.~W.~Nam, D.~Oblak, W.~Tittel, Physics Review Letters \textbf{115}, 140501 (2015).

\bibitem{MacFarlane1987}
R.~Macfarlane and R.~Shelby, Journal of Luminescence \textbf{36}, 179 (1987).

\bibitem{Johnson1988}
L.~M.~Johnson and C.~H.~Cox, Journal of Lightwave Technology \textbf{6}, 109 (1988).
  
\bibitem{in-prep}
L.~Veissier \textit{et al.}, in preparation.

\bibitem{hayes_non-photochemical_1978}
J.~M. Hayes and G.~J. Small, Chemical Physics \textbf{27}, 151 (1978).

\bibitem{bottger_spectroscopy_2006}
T.~B\"{o}ttger, Y.~Sun, C.~W. Thiel, and R.~L. Cone, Physical Review B \textbf{74}, 075107 (2006).

\bibitem{Hastings2008}
S.~R. Hastings-Simon, B.~Lauritzen, M.~U. Staudt, J.~L.~M. van Mechelen, C.~Simon, H.~de~Riedmatten, M.~Afzelius, and N.~Gisin, Physical Review B \textbf{78}, 085410 (2008).

\bibitem{Bottger2006}
T.~B\"ottger, C.~W. Thiel, Y.~Sun, and R.~L. Cone, Phys. Rev. B \textbf{73}, 075101 (2006).
  
\bibitem{Macfarlane2006}
R.~M.~Macfarlane, Y.~Sun, P.~B.~Sellin, and R.~L.~Cone, Phys. Rev. Lett. \textbf{96}, 033602 (2006).

\bibitem{Staudt2006720}
M.~U. Staudt, S.~R. Hastings-Simon, M.~Afzelius, D.~Jaccard, W.~Tittel, and N.~Gisin, Optics Communications \textbf{266}, 720  (2006).

\bibitem{Kurkin1980}
I.~Kurkin and K.~Chernov, Physica B+C \textbf{101}, 233  (1980).

\bibitem{bottger_laser_2002}
T.~B\"{o}ttger, \enquote{Laser frequency stabilization to spectral hole burning frequency references in erbium-doped crystals: {Material} and device optimization}, Ph.D. thesis, Montana State University (2002).

\bibitem{milori_optical_1995}
D.~M. B.~P. Milori, I.~J. Moraes, A.~C. Hernandes, R.~R. de~Souza, M.~S. Li, M.~C. Terrile, and G.~E. Barberis, Physical Review B \textbf{51}, 3206 (1995).
  
\bibitem{ball_low-temperature_1961}
M.~Ball, G.~Garton, M.~J.~M. Leask, D.~Ryan, and W.~P. Wolf, Journal of Applied Physics \textbf{32}, S267 (1961).

\bibitem{buchenau_low-frequency_1986}
U.~Buchenau, M.~Prager, N.~N\"ucker, A.~J. Dianoux, N.~Ahmad, and W.~A. Phillips, Physical Review B \textbf{34}, 5665 (1986).

\bibitem{Orbach1961}
R.~Orbach, Proceedings of the Royal Society of London. Series A. Mathematical and Physical Sciences \textbf{264}, 458 (1961).
  
\bibitem{fraval_dynamic_2005}
E.~Fraval, M.~J. Sellars, and J.~J. Longdell, Physical Review Letters \textbf{95}, 030506 (2005).

\bibitem{probst_anisotropic_2013}
S.~Probst, H.~Rotzinger, S.~W\"unsch, P.~Jung, M.~Jerger, M.~Siegel, A.~V. Ustinov, and P.~A. Bushev, Physical Review Letters \textbf{110}, 157001 (2013).
  
\bibitem{jobez_coherent_2015}
P.~Jobez, C.~Laplane, N.~Timoney, N.~Gisin, A.~Ferrier, P.~Goldner, and M.~Afzelius, Physical Review Letters \textbf{114}, 230502 (2015).

\bibitem{Haworth2010}
R.~Haworth, G.~Mountjoy, M.~Corno, P.~Ugliengo, and R.~J. Newport, Phys. Rev. B \textbf{81}, 060301 (2010).
  
\bibitem{sun_exceptionally_2006}
Y.~Sun, R.~L. Cone, L.~Bigot, and B.~Jacquier, Optics Letters \textbf{31}, 3453 (2006).

\bibitem{meltzer_electron-phonon_2000}
R.~S. Meltzer and K.~S. Hong, Physical Review B \textbf{61}, 3396 (2000).


\end{thebibliography}

\end{document}